\documentclass[aps,prb,twocolumn,superscriptaddress,showpacs,floatfix,amsmath,amssymb]{revtex4}

\usepackage{graphicx}
\usepackage{dcolumn}
\usepackage{bm}
\usepackage[below]{placeins}

\begin{document}

\title{Lattice distortion and stripe-like antiferromagnetic order in Ca$_{10}$(Pt$_{3}$As$_{8}$)(Fe$_{2}$As$_{2}$)$_{5}$}

\author{A.~Sapkota}
\affiliation{Ames Laboratory, U.\,S.\,DOE, Ames, Iowa 50011, USA} 
\affiliation{Department of Physics and Astronomy, Iowa State 
University, Ames, Iowa 50011, USA}

\author{G.\,S.~Tucker}
\affiliation{Ames Laboratory, U.\,S.\,DOE, Ames, Iowa 50011, USA} 
\affiliation{Department of Physics and Astronomy, Iowa State 
University, Ames, Iowa 50011, USA}

\author{M.~Ramazanoglu}
\affiliation{Ames Laboratory, U.\,S.\,DOE, Ames, Iowa 50011, USA} 
\affiliation{Department of Physics and Astronomy, Iowa State 
University, Ames, Iowa 50011, USA}

\author{W.~Tian}
\affiliation{Quantum Condensed Matter Division, Oak Ridge 
National Laboratory, Oak Ridge, Tennessee 37831, USA} 

\author{N.~Ni}
\affiliation{Department of Physics and Astronomy, University of 
California, Los Angeles, California 90095, USA} 
\affiliation{Department of Chemistry, Princeton University, 
Princeton, New Jersey 08544, USA}

\author{R.\,J.~Cava}
\affiliation{Department of Chemistry, Princeton University, 
Princeton, New Jersey 08544, USA}

\author{R.\,J.~McQueeney}
\affiliation{Ames Laboratory, U.\,S.\,DOE, Ames, Iowa 50011, USA} 
\affiliation{Department of Physics and Astronomy, Iowa State 
University, Ames, Iowa 50011, USA} 
\affiliation{Oak Ridge 
National Laboratory, Oak Ridge, Tennessee 37831, USA} 

\author{A.\,I.~Goldman}
\affiliation{Ames Laboratory, U.\,S.\,DOE, Ames, Iowa 50011, USA} 
\affiliation{Department of Physics and Astronomy, Iowa State 
University, Ames, Iowa 50011, USA}

\author{A.~Kreyssig}
\affiliation{Ames Laboratory, U.\,S.\,DOE, Ames, Iowa 50011, USA} 
\affiliation{Department of Physics and Astronomy, Iowa State 
University, Ames, Iowa 50011, USA}

\date{\today}

\begin{abstract}

Ca$_{10}$(Pt$_{3}$As$_{8}$)(Fe$_{2}$As$_{2}$)$_{5}$ is the parent 
compound for a class of Fe-based high-temperature superconductors 
where superconductivity with transition temperatures up to 30\,K 
can be introduced by partial element substitution. We present a 
combined high-resolution high-energy x-ray diffraction and 
elastic neutron scattering study on a 
Ca$_{10}$(Pt$_{3}$As$_{8}$)(Fe$_{2}$As$_{2}$)$_{5}$ single 
crystal. This study reveals the microscopic nature of two 
distinct and continuous phase transitions to be very similar to 
other Fe-based high-temperature superconductors: an orthorhombic 
distortion of the high-temperature tetragonal Fe-As lattice below 
$T_{\textrm{S}}$\,=\,110(2)\,K followed by stripe-like 
antiferromagnetic ordering of the Fe moments below 
$T_{\textrm{N}}$\,=\,96(2)\,K. These findings demonstrate that 
major features of the Fe-based high-temperature superconductors 
are very robust against variations in chemical constitution as 
well as structural imperfection of the layers separating the 
Fe-As layers from each other and confirms that the Fe-As layers 
primarily determine the physics in this class of material.

\end{abstract}

\pacs{74.70.Xa, 75.25.-j, 61.50.Ks, 75.30.Kz}

\maketitle

In early 2011, a new class of Fe-based high-temperature 
superconductors was discovered, the 
Ca$_{10}$(Pt$_{3}$As$_{8}$)(Fe$_{2-x}$Pt$_{x}$As$_{2}$)$_{5}$ 
(10-3-8) and 
Ca$_{10}$(Pt$_{4}$As$_{8}$)(Fe$_{2-x}$Pt$_{x}$As$_{2}$)$_{5}$ 
(10-4-8) compounds.\cite{Kakiya-2011,Ni-2011,Loehnert-2011} The 
parent 10-4-8 compound shows superconductivity with a transition 
temperature 
$T_{\textrm{c}}$\,$\sim$\,38\,K\cite{Kakiya-2011,Stuerzer-2012} 
which can be suppressed by applied pressure or chemical 
doping.\cite{Stuerzer-2012,Nohara-2012,Tamegai-2013} In the 
partner system, the parent 10-3-8 compound is 
non-superconducting. However, superconductivity can be induced by 
partial element substitution of Fe by Pt with $T_{\textrm{c}}$ 
values up to 
14\,K,\cite{Kakiya-2011,Ni-2011,Loehnert-2011,Stuerzer-2012,Tamegai-2013,Cho-2012,Xiang-2012,Ding-2012} 
by partial substitution of Ca by La with maximum $T_{\textrm{c}}$ 
of 30\,K,\cite{Stuerzer-2012,Ni-2013,Kim-2013} or by applied 
pressure.\cite{Gao-2014} Varying the Pt concentration in either 
compound yields charge doping in the 
(Fe$_{2-x}$Pt$_{x}$As$_{2}$)$_{5}$ (Fe-As) 
layers\cite{Ni-2011,Loehnert-2011,Kobayashi-2012} and in the 
Pt$_{3+y}$As$_{8}$ or Pt$_{4+y}$As$_{8}$ (Pt-As) 
layers.\cite{Loehnert-2011,Stuerzer-2012} Recently, the 10-3-8 
and 10-4-8 class expanded with the discovery of 
(Ca$_{1-x}$La$_{x}$)$_{10}$(Pd$_{3}$As$_{8}$)(Fe$_{2-y}$Pd$_{y}$As$_{2}$)$_{5}$ 
and Ca$_{10}$(Ir$_{4}$As$_{8}$)(Fe$_{2}$As$_{2}$)$_{5}$ compounds 
with $T_{\textrm{c}}$ values up to 17\,K and 16\,K, respectively. 
Their crystal and electronic structures are almost identical to 
their 10-3-8 and 10-4-8 elder sibling 
compounds.\cite{Hieke-2013,Kudo-2013}

The crystal structures of both 10-3-8 and 10-4-8 compounds share 
a common feature with other families of Fe-based high-temperature 
superconductors: a stacking of Fe-As layers with equivalent 
arrangement and electronic configuration to that in the parent 
BaFe$_{2}$As$_{2}$ and LaFeAsO 
compounds.\cite{Nohara-2012,Ni-2011,Kakiya-2011,Loehnert-2011,Stuerzer-2012} 
However, there are several distinguishing structural features of 
the 10-3-8 and 10-4-8 compounds: (i) the Fe-As layers are 
separated by slightly puckered Pt-As layers and Ca planes, 
yielding a much larger separation of the Fe-As layers, 
approximately 10.3\,\AA; (ii) the arrangement of atoms in the 
Pt-As layers leads to chemical structures with lower symmetries 
than found in other Fe-based high-temperature superconductors, 
triclinic $P\,\overline{1}$ for the 10-3-8 compound, and 
tetragonal $P\,4/n$, monoclinic $P\,2_{1}/n$, or triclinic 
$P\,\overline{1}$ for the 10-4-8 compound in different structure 
studies;\cite{Ni-2011,Kakiya-2011,Loehnert-2011,Stuerzer-2012} 
(iii) imperfections in the chemical structure have been observed 
by diffuse scattering and streaking of spots in x-ray diffraction 
studies\cite{Ni-2011,Loehnert-2011} that are interpreted as 
resulting mainly from stacking disorder of the Pt-As layers. 

Beyond these structural differences, the physical properties of 
10-3-8 and 10-4-8 compounds are very reminiscent of other 
Fe-based high-temperature superconductors. For example, 
angle-resolved photoemission studies supported by band-structure 
and phonon-spectra calculations have shown that the electronic 
structure is quite similar to that observed for 
BaFe$_{2}$As$_{2}$.\cite{Neupane-2012,Nakamura-2013,Shen-2013,Thirupathaiah-2013,Berlijn-2013} 
Inelastic neutron scattering measurements on large single 
crystals\cite{Sato-2012,Ikeuchi-2013} demonstrated that magnetic 
fluctuations occur at the same stripe-like antiferromagnetic wave 
vector as observed in other Fe-based high-temperature 
superconductors. However, no indication of magnetic order could 
be found -- likely related to the high Pt 
content\cite{Sato-2012,Ikeuchi-2013} and the reported 
inhomogeneity in one of the samples.\cite{Ikeuchi-2013} 
Furthermore, in resistance measurements on 
Ca$_{10}$(Pt$_{3}$As$_{8}$)(Fe$_{2}$As$_{2}$)$_{5}$ single 
crystals, Ni $et\,al.$\cite{Ni-2013} and Gao 
$et\,al.$\cite{Gao-2014} found two features that they interpreted 
as a structural transition at $T_{1}$\,=\,103\,K and a magnetic 
transition at $T_{2}$\,=\,95\,K. The presence of a structural 
transition is consistent with the observation of domain patterns 
in polarized-light images due to a lowering of the lattice 
symmetry at low temperatures,\cite{Cho-2012} and recent 
high-resolution x-ray diffraction experiments on powder 
samples.\cite{Stuerzer-2013} Susceptibility measurements on a 
powder sample,\cite{Nohara-2012} muon spin-resonance ($\mu$SR) 
measurements on powder samples,\cite{Stuerzer-2013} and $^{75}$As 
nuclear magnetic resonance (NMR) measurements on a single 
crystal\cite{Zhou-2013} showed features interpreted as an onset 
of antiferromagnetic order below 120\,K for powder samples and 
below 100\,K for single crystals, respectively. Nevertheless, 
there are indications of a strong sensitivity of physical 
properties to different sample preparation routes, $e.\,g.$, in 
specific heat measurements. Whereas the observed jump at 
$T_{\textrm{c}}$ is much smaller for superconducting 
polycrystalline material\cite{Kim-2013} than one would expect 
from the almost universal behavior in Fe-based high-temperature 
superconductors,\cite{Budko-2009} it follows the anticipated 
behavior much more closely for single crystals.\cite{Ni-2013} 

Previous microscopic studies have been performed, for the most 
part, on different samples and it is difficult to compare and 
correlate the structural and magnetic properties due to variances 
in sample preparation and the apparent wide spread of 
properties.\cite{Ni-2011,Kakiya-2011,Loehnert-2011,Stuerzer-2012,Nohara-2012,Stuerzer-2013} 
Therefore, important questions concerning the properties of 
10-3-8 and 10-4-8 compounds remain unanswered. Which phase 
transitions are intrinsic to homogeneous single-phase samples and 
what is their nature? In particular, are the structural and 
magnetic transitions similar to lattice distortions and 
antiferromagnetic ordering observed in other Fe-based 
high-temperature superconductors? Are these phase transitions 
coupled and discontinuous (1$^{\textrm{st}}$ order) or continuous 
(2$^{\textrm{nd}}$ order)? 

Here we present a combined high-resolution high-energy x-ray 
diffraction and elastic neutron scattering study to answer most 
of the aforementioned questions on a microscopic level in an 
unambiguous manner by performing  measurements on the same, 
homogenous Ca$_{10}$(Pt$_{3}$As$_{8}$)(Fe$_{2}$As$_{2}$)$_{5}$ 
single crystal. We find two distinct and continuous phase 
transitions: an orthorhombic distortion of the high-temperature 
square-like Fe lattice followed by a stripe-like 
antiferromagnetic ordering of the Fe moments at slightly lower 
temperature.  

A Ca$_{10}$(Pt$_{3}$As$_{8}$)(Fe$_{2}$As$_{2}$)$_{5}$ single 
crystal of dimension 2$\times$2$\times$0.2\,mm$^{3}$ was prepared 
by solution growth\cite{Ni-2011} and was characterized as 
described in Ref.\,\onlinecite{Cho-2012}. Herein, we disregard 
the very small partial element substitution of 0.4\% Pt for Fe 
for similarly prepared samples.\cite{Cho-2012,Zhou-2013,Gao-2014} 
The high-resolution high-energy x-ray diffraction study was 
performed at station 6-ID-D at the Advanced Photon Source, 
Argonne National Laboratory. The use of x-rays with an energy of 
100.5\,keV minimizes sample absorption and allows us to probe the 
entire bulk of the sample. The sample was mounted on a Cu post in 
the center of a Be sample can filled with He exchange gas, a Be 
heat shield, and a Be vacuum shroud, all in an APD He closed 
cycle cryostat. Extended regions of selected reciprocal lattice 
planes were recorded by a MAR345 image plate system positioned 
1497\,mm behind the sample as the sample was rocked through two 
independent angles up to $\pm$3.2$^{\circ}$ about axes 
perpendicular to the incident beam with a size of 
0.1$\times$0.1\,mm$^{2}$.\cite{Kreyssig-2007} High-resolution 
images were measured by employing a ScintX area detector 
positioned 2713\,mm behind the sample while rocking one of these 
angles after aligning the selected Bragg peak of interest. The 
elastic neutron-scattering measurements were performed on the 
HB-1A fixed incident energy (14.6\,meV) triple-axis spectrometer 
at the High Flux Isotope Reactor, Oak Ridge National Laboratory 
using two pyrolytic graphite filters and 48'-48'-40'-80' 
collimation. The same sample employed in the x-ray diffraction 
study was wrapped in Al foil and mounted on an Al sheet at the 
center of an Al sample can filled with He exchange gas, an Al 
heat shield, and an Al vacuum shroud, all in a He closed cycle 
cryostat. The ($H$\,$H$\,$L$) plane was selected as scattering 
plane in the body-centered tetragonal pseudo-symmetric coordinate 
system defined below. 

\begin{figure*}
\centering\includegraphics[width=0.95\linewidth]{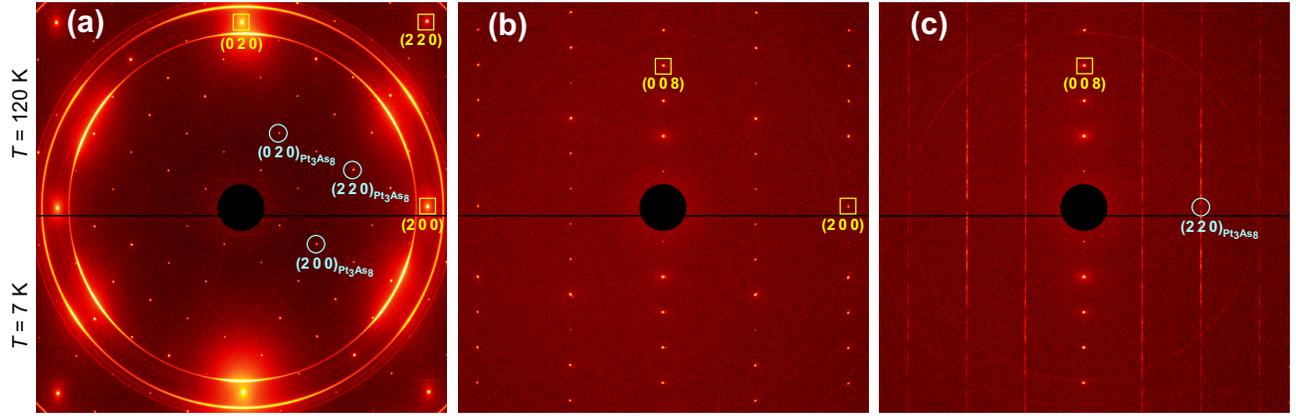} 
\caption{\label{fig:fighxrdoverview}(Color online) High-energy 
x-ray diffraction patterns of the 
Ca$_{10}$(Pt$_{3}$As$_{8}$)(Fe$_{2}$As$_{2}$)$_{5}$ single 
crystal measured by the method described in the text. Reciprocal 
lattice plane recorded (a) with the incident x-ray beam 
perpendicular to the plate-like crystals, (b) after rotating the 
sample by -90$^{\circ}$ around the horizontal axis in (a), and 
(c) after rotating the sample by -18.44$^{\circ}$ around the 
vertical axis in (b). The lower halves of the patterns were 
measured at $T$\,=\,7\,K and the upper halves at $T$\,=\,120\,K. 
Selected Bragg peaks related to the body-centered tetragonal base 
structure built by the Fe-As layers and Ca planes are marked with 
squares. Circles depict selected Bragg peaks related to the 
Pt$_{3}$As$_{8}$ superstructure. The ring-like scattering 
features are caused by the Cu-sample holder and the Be domes.}
\end{figure*}

Figure 1(a) shows the high-energy x-ray diffraction pattern of 
the ($H$\,$K$\,0) reciprocal lattice plane recorded at 
$T$\,=\,120\,K well above all reported phase transition 
temperatures. Two families of resolution-limited Bragg peaks of 
the Ca$_{10}$(Pt$_{3}$As$_{8}$)(Fe$_{2}$As$_{2}$)$_{5}$ single 
crystal have been observed: (i) strong Bragg peaks on a large 
square-like reciprocal lattice related to the dimensions and 
geometry of the Fe-As layers, and (ii) weaker Bragg peaks with a 
smaller square-like reciprocal lattice related to the complex 
Pt$_{3}$As$_{8}$ layer structure with its in-plane triclinic unit 
cell that is essentially a $\sqrt{5}\times\sqrt{5}$ superlattice 
forming along the [2\,1\,0] direction of the tetragonal cell 
defined by the Fe-As layers. This results in an in-plane 
inclusion angle of arctan$\frac{1}{2}$\,=\, 26.56$^{\circ}$ and a 
length relation $a$\,=$\sqrt{5}$$a_{0}$ between the two 
lattices.\cite{Ni-2011,Neupane-2012} We employ the description of 
the chemical structure and the reciprocal space in terms of a 
body-centered tetragonal unit cell common to other Fe-based 
high-temperature superconductors with lattice parameters 
$a_{0}$\,=\,3.912\,\AA~ and $c_{0}$\,=\,20.58\,\AA~ related to 
the base structure built by the Fe-As layers and Ca planes as 
illustrated in Fig.\,1 in Ref.\,\onlinecite{Neupane-2012}. This 
structure and lattice yield the strong Bragg peaks depicted in 
(i). The additional weaker Bragg peaks (ii) in 
Fig.\,\ref{fig:fighxrdoverview}(a) arise from the complex 
Pt$_{3}$As$_{8}$ layer structure and will be described as 
superstructure hereafter. We note that only one of the possible 
two orientations of the Pt$_{3}$As$_{8}$ superstructure relative 
to the Fe-As base structure is observed for the entire sample. 

The diffraction pattern of the ($H$\,0\,$L$) reciprocal lattice 
plane shown in Fig.\,\ref{fig:fighxrdoverview}(b) was recorded in 
the same manner as that in Fig.\,\ref{fig:fighxrdoverview}(a) 
after rotating the sample by -90$^{\circ}$ around the [1\,0\,0] 
direction.  Only resolution limited Bragg peaks related to the 
body-centered tetragonal unit cell are observed. This is in 
contrast to the diffraction pattern shown in 
Fig.\,\ref{fig:fighxrdoverview}(c) where the sample was rotated 
such that the [0\,0\,1] direction was still vertical and the 
[2\,2\,0]$_{\textrm{Pt}_{3}\textrm{As}_{8}}$ marked direction in 
Fig.\,\ref{fig:fighxrdoverview}(a) was horizontal. The strong 
resolution limited Bragg peaks along the [0\,0\,1] direction are 
associated with the base structure. The additional streak-like 
diffraction features arise from the Pt$_{3}$As$_{8}$ 
superstructure, indicating significant disorder along the 
\textbf{c} direction in contrast to a well-ordered structure in 
the (\textbf{a}\,\textbf{b}) plane. This disorder appears to be 
static as demonstrated by very similar features in the 
measurements at $T$\,=\,120\,K in the upper halves of 
Fig.\,\ref{fig:fighxrdoverview} and at $T$\,=\,7\,K in the lower 
halves of Fig.\,\ref{fig:fighxrdoverview}. Additional 
measurements at $T$\,=\,20\,K, 40\,K, 60\,K, 80\,K, and 100\,K 
yield similar features in the diffraction patterns and support 
this interpretation. 

\begin{figure}
\centering\includegraphics[width=0.93\linewidth]{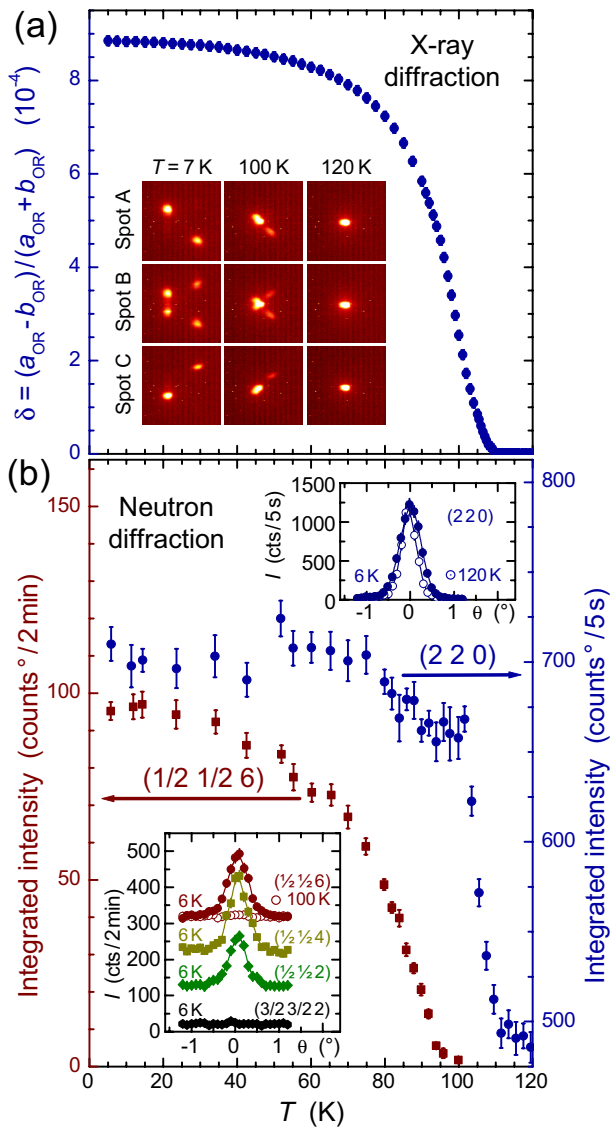} 
\caption{\label{fig:figtdependence}(Color online) Temperature 
dependence of the order parameters of the 
Ca$_{10}$(Pt$_{3}$As$_{8}$)(Fe$_{2}$As$_{2}$)$_{5}$ single 
crystal. (a) Lattice distortion extracted from the 
high-resolution high-energy x-ray diffraction data shown in the 
inset measured at three different spots A-C on the sample in the 
surrounding of the (2\,2\,0) Bragg peak of the tetragonal 
body-centered base structure. The origin of the reciprocal space 
and the direct-beam direction is left from and far out of the 
shown detector area. (b) Integrated intensities of the 
antiferromagnetic ($\frac{1}{2}$\,$\frac{1}{2}$\,6) and nuclear 
(2\,2\,0) Bragg peaks measured in the neutron diffraction study 
of the Ca$_{10}$(Pt$_{3}$As$_{8}$)(Fe$_{2}$As$_{2}$)$_{5}$ single 
crystal. The insets show rocking scans through selected 
antiferromagnetic and nuclear Bragg peaks presented with 
intensity offsets for clarity.}
\end{figure}

However, qualitative changes are observed in the ($H$\,$K$\,0) 
plane diffraction patterns measured at $T$\,=\,100\,K and below; 
in particular, the Bragg peaks split or show significant 
broadening observable in the high-resolution patterns shown in 
the inset in Fig.\,\ref{fig:figtdependence}(a) measured at 
different positions on the sample. For the spots labeled A and C, 
a splitting of the single high-temperature (2\,2\,0) Bragg peak 
into a pair of peaks is observed at $T$\,=\,100\,K, which evolves 
in separation and shape down to $T$\,=\,7\,K. The (2\,2\,0) Bragg 
peak splits into four peaks at spot B which can be explained in 
terms of the tetragonal-to-orthorhombic distortion of the base 
structure and the arising domain structure previously reported 
for BaFe$_{2}$As$_{2}$ and 
GdFeAsO.\cite{Tanatar-2009,Nitsche-2013} This is in excellent 
agreement with the analysis of similar diffraction patterns in 
these 
compounds.\cite{Tanatar-2009,Tanatar-2010,Blomberg-2012,Nitsche-2013} 
The diffraction patterns at spots A-C are similar to spots A-C 
shown in Figs.\,2 and 3 in Ref.\,\onlinecite{Tanatar-2009} 
considering the 45$^{\circ}$ rotated detector orientation between 
the two measurements. This demonstrates a similar order of 
magnitude for the size of the orthorhombic distorted domains in 
Ca$_{10}$(Pt$_{3}$As$_{8}$)(Fe$_{2}$As$_{2}$)$_{5}$ as in 
BaFe$_{2}$As$_{2}$, in agreement with the similarity in polarized 
light-microscopic pictures observed in both 
compounds.\cite{Cho-2012,Tanatar-2009} The detailed temperature 
dependence is similar for all three sample spots A-C in 
Ca$_{10}$(Pt$_{3}$As$_{8}$)(Fe$_{2}$As$_{2}$)$_{5}$, 
demonstrating a high degree of homogeneity across the entire 
sample.

Based on the occurrence of a lattice distortion in 
Ca$_{10}$(Pt$_{3}$As$_{8}$)(Fe$_{2}$As$_{2}$)$_{5}$ similar to 
that observed in BaFe$_{2}$As$_{2}$ and GdFeAsO, the question 
arises as to whether the magnetic order of the Fe moments is also 
similar, $i.\,e.$, a stripe-like arrangement in the 
(\textbf{a}\,\textbf{b}) plane. Indeed, magnetic Bragg peaks 
could be found at several positions 
($\frac{u}{2}$\,$\frac{v}{2}$\,$w$) with $u$ and $v$\,=\,odd and 
$w$\,=\,even in our neutron diffraction study as shown in the 
left inset in Fig.\,\ref{fig:figtdependence}(b). An intensive 
search for magnetic Bragg peaks at positions 
($\frac{u}{2}$\,$\frac{v}{2}$\,$w$) with $u$, $v$, and 
$w$\,=\,odd performed around ($\frac{1}{2}$\,$\frac{1}{2}$\,1), 
($\frac{1}{2}$\,$\frac{1}{2}$\,3), 
($\frac{1}{2}$\,$\frac{1}{2}$\,5) , and 
($\frac{1}{2}$\,$\frac{1}{2}$\,7) failed to observe any Bragg 
peaks above a background level of 33 counts in 3\,min at 
$T$\,=\,6\,K. The observation of magnetic Bragg peaks related to 
the propagation vector ($\frac{1}{2}$\,$\frac{1}{2}$\,0) and the 
absence of peaks related to the propagation vector 
($\frac{1}{2}$\,$\frac{1}{2}$\,1) points to an antiferromagnetic 
arrangement of the Fe moments along the [1\,1\,0] direction and 
ferromagnetic arrangements in both perpendicular directions. A 
similar stripe-like ordering in the Fe layers with ferromagnetic 
arrangement between the layers was observed, $e.\,g.$, in 
CeFeAsO, PrFeAsO, and 
NdFeAsO.\cite{Zhao-2008a,Zhao-2008b,Tian-2010} The 
antiferromagnetic Bragg peaks shown in the left inset in  
Fig.\,\ref{fig:figtdependence}(b) are very weak when the 
scattering vector direction is close to the [1\,1\,0] direction 
and become stronger with increasing angle between both directions 
despite Lorentz-factor influence and potential structure factor 
effects. In addition, no Bragg peak could be found above 
background at the ($\frac{1}{2}$\,$\frac{1}{2}$\,0) position. 
Both observations together demonstrate that the 
antiferromagnetically ordered Fe moments are primarily 
collinearly aligned along the [1\,1\,0] direction. From the 
similar shape and half width of the magnetic and nuclear Bragg 
peaks as shown in the insets in Fig.\,\ref{fig:figtdependence}(b) 
we conclude that the antiferromagnetic order is long-range in 
nature. 

In the following discussion, the temperature dependence of both 
ordering phenomena is elucidated. The value of $\delta = 
(a_{\textrm{OR}}-b_{\textrm{OR}})/(a_{\textrm{OR}}+b_{\textrm{OR}})$ 
is used as the order parameter for the observed orthorhombic 
lattice distortion.\cite{Nandi-2010} It is determined by 
extracting the in-plane orthorhombic lattice parameters 
$a_{\textrm{OR}}$ and $b_{\textrm{OR}}$ from the longitudinal 
splitting of the x-ray Bragg peaks near the (2\,2\,0) peak. The 
lattice distortion sets in below $T_{\textrm{S}}$\,=\,110(2)\,K 
as shown in Fig.\,\ref{fig:figtdependence}(a) and reaches a 
maximum value of $\delta$\,=\,8.8(2)$\times$10$^{-4}$, the same 
order of magnitude as observed in other Fe-based high-temperature 
superconductors.\cite{Nandi-2010} The lattice distortion causes 
an increase of the nuclear (2\,2\,0) Bragg peak in the neutron 
diffraction study through an extinction 
release\cite{Kreyssig-2010} below $T_{\textrm{S}}$ as illustrated 
in Fig.\,\ref{fig:figtdependence}(b), confirming consistent 
temperature scales for both scattering experiments performed on 
the same sample. The antiferromagnetic order sets in at 
$T_{\textrm{N}}$\,=\,96(2)\,K well below $T_{\textrm{S}}$ as 
demonstrated by the temperature-dependent intensity of the 
antiferromagnetic ($\frac{1}{2}$\,$\frac{1}{2}$\,6) Bragg peak 
shown in Fig.\,\ref{fig:figtdependence}(b). The temperature 
dependence is consistent with a continuous (or 2$^{\textrm{nd}}$ 
order) phase transition at $T_{\textrm{N}}$ similar to the 
structural phase transition at $T_{\textrm{S}}$ characterized by 
the lattice distortion $\delta$ shown in 
Fig.\,\ref{fig:figtdependence}(a). This is also consistent with 
the lack of a step-like feature in the temperature-dependent 
curve of the distortion $\delta$ around $T_{\textrm{N}}$ as 
observed in BaFe$_{2}$As$_{2}$ and derived 
compounds.\cite{Kim-2011} 

Summarizing, Ca$_{10}$(Pt$_{3}$As$_{8}$)(Fe$_{2}$As$_{2}$)$_{5}$ 
is very similar to other Fe-based high-temperature 
superconductors not only in macroscopic physical properties, like 
the reported occurrence of superconductivity by partial element 
substitution of Fe by Pt, but also in its microscopic physical 
properties: Below $T_{\textrm{S}}$\,=\,110(2)\,K, the 
high-temperature square-like iron lattice gets distorted along 
its diagonal directions; below $T_{\textrm{N}}$\,=\,96(2)\,K, the 
Fe moments order in a long-range stripe-like antiferromagnetic 
structure built by Fe moments aligned and antiferromagnetically  
correlated along one of these directions and with ferromagnetic 
correlations perpendicular to it. Both transitions are continuous 
and well separated from each other. This similarity in physical 
properties is remarkable based on the observed strong differences 
in the chemical order: the Fe-As layers are much further 
separated from each other by Pt-As layers and neighboring Ca 
planes yielding a much more complex structure with lower crystal 
symmetry and severe chemical disorder in the sequence of the 
Pt-As layers. These intermediate layers modify the physical 
properties, $e.\,g.$, yielding only weak ferromagnetic 
interactions in the direction perpendicular to the 
layers;\cite{Shen-2013,Berlijn-2013} however, major physical 
features seem very robust against variations of these 
intermediate layers in their chemical constitution as well as 
their structural perfection. These findings suggest that the 
Fe-As layer in tetragonal arrangement are key in determining the 
major physical properties of the class 
of Fe-based high-temperature superconductors.\\
\\
The authors appreciate the help in sample preparation by 
M.\,G.~Kim and W.~Jayasekara, and the excellent support of the 
x-ray diffraction measurements by D.\,S.~Robinson. Work at Ames 
Laboratory was supported by the U.\,S. Department of Energy, 
Office of Basic Energy Science, Division of Materials Sciences 
and Engineering. Ames Laboratory is operated for the U.\,S. 
Department of Energy by Iowa State University under Contract 
No.~DE-AC02-07CH11358. The research conducted at Argonne National 
Laboratory and Oak Ridge National Laboratory was sponsored by the 
U.\,S. Department of Energy, Office of Basic Energy Science, 
Scientific User Facilities Division. Work at Princeton University 
is supported by the Air Force Office of Scientific Research, 
Multidisciplinary University Research Initiative on 
Superconductivity.

\end{document}